\documentclass[journal]{IEEEtran}
\usepackage{amssymb}
\usepackage{pifont}
\usepackage{bbding}

\usepackage{amssymb, graphicx, subfigure}
\usepackage[ruled, lined, linesnumbered, commentsnumbered]{algorithm2e}
\usepackage{cleveref}
\usepackage[cmex10]{amsmath}

\usepackage{array, setspace}
\usepackage{mdwmath}
\usepackage{mdwtab}
\usepackage{multirow}
\usepackage{color}

\usepackage[pagewise]{lineno}
\usepackage{tikz}
\usetikzlibrary{shapes,snakes}

\title{A Novel Compressive Sensing based Enhanced Multiplexing Scheme for MIMO System}
\IEEEaftertitletext{\vspace{-1\baselineskip}}
\author{Chanzi~Liu, %
        Qingchun~Chen, Xiaohu~Tang %
}

\begin{document}
\maketitle
\begin{abstract}
 A novel compressive-sensing based signal multiplexing scheme is proposed in this paper to further improve the multiplexing gain for multiple input multiple output (MIMO) system. At the transmitter side, a Gaussian random measurement matrix in compressive sensing is employed before the traditional spatial multiplexing in order to carry more data streams on the available spatial multiplexing streams of the underlying MIMO system. At the receiver side, it is proposed to reformulate the detection of the multiplexing signal into two steps. In the first step, the traditional MIMO equalization can be used to restore the transmitted spatial multiplexing signal of the MIMO system. While in the second step, the standard optimization based detection algorithm assumed in the compressive sensing framework is utilized to restore the CS multiplexing data streams, wherein the exhaustive over-complete dictionary is used to guarantee the sparse representation of the CS multiplexing signal. In order to avoid the excessive complexity, the sub-block based dictionary and the sub-block based CS restoration is proposed. Finally, simulation results are presented to show the feasibility of the proposed CS based enhanced MIMO multiplexing scheme. And our efforts in this paper shed some lights on the great potential in further improving the spatial multiplexing gain for the MIMO system.
\end{abstract}%
\begin{keywords}
Compressive sensing, Multiplexing, Over-complete dictionary.
\end{keywords}%
\section{Introduction}
\IEEEPARstart{H}{ow} to fulfill the ever increasing data traffic always impose critical challenge for the wireless communication design. Due to the scarce radio spectrum resource, it is imperative to use spatial multiplexing gain of multiple input multiple output (MIMO) system without extra bandwidth expansion requirements. However, the spatial multiplexing gain is limited by the smaller one of both transmit antenna number and receive antenna number, which makes the huge antenna arrays seem to be one of the promising candidate in future wireless communications, for instance the fifth-generation (5G) wireless communications systems \cite{Demestichas13}. However, it is still a challenging issue to support a huge antenna array due to the physical size limit, the required realization complexity for wireless communication system. How to further improve the multiplexing gain for the given MIMO system is thus an important problem. And this is exactly the motivation of our work in this paper.

Compressive Sensing (CS) \cite{Donoho06A}-\cite{Baraniuk07} is a novel technique that enables efficient sampling below Nyquist rate, without (or with little) sacrificing reconstruction quality. In this paper, we propose to integrate this novel signal processing strategy into the signal multiplexing scheme for MIMO system to further improve its multiplexing gain. In order to do that, a Gaussian random measurement matrix in compressive sensing is employed at the transmitter side before the traditional MIMO spatial multiplexing. Because of the dimensional suppression through the proposed measurement processing, now we may transform more data streams to less data streams, which matches the given spatial multiplexing streams of the underlying MIMO system. While at the receiver side, the detection of the multiplexing signal will be accomplished via two steps. More specifically, the traditional MIMO equalization will be performed in the first step to restore the transmitted spatial multiplexing signal of the MIMO system. While in the second step, it is proposed to follow the standard optimization based detection assumed in the compressive sensing framework to restore the CS multiplexing data streams. According to the CS principle, the sparse representation of the CS-based multiplexing signal will be one of the essential problem to guarantee the restoration success rate. Unlike the traditional analog signal, now each individual entry of the CS-based multiplexing signal is discrete digital modulation signal. For the sake of restoration requirement, the exhaustive over-complete dictionary is used to secure the sparse representation of the CS multiplexing signal. However, this may cause an excessive complexity, especially when either the modulation alphabet size or the number of data streams is large. In order to make the calculation complexity affordable, the sub-block based dictionary and the sub-block based CS restoration is proposed. Simulation results are presented to show the feasibility of the proposed CS based enhanced MIMO multiplexing scheme. And our analysis in this paper sheds some lights on the potential in further improving the spatial multiplexing gain for the MIMO system.

The remainder of this paper is organized as follows. The basic idea of CS and problem reformulation will be introduced in Section II. Nextly, we deduce the implement of CS-based MIMO detection and practical multiplexing algorithm design in Section III. Experimental results are demonstrated in Section IV. Finally, we conclude this paper in Section V.

\section{A Brief Review of CS and MIMO Technology}
\subsection{The Basics of CS}
CS converts the high-dimensional signal into a significantly lower dimensional measurement space. For example, the real-valued discrete signal $\boldsymbol{x}=[x_1, x_2, \cdots, x_L]$, with sparsity $\|\boldsymbol{x}\|_0=K$, where $\|\cdot\|_0$ denotes the $\ell_0$-norm counting the number of nonzero items, can be transformed into an $M \times 1$ measurement vector by taking $M$ linear, nonadaptive measurement as below
\begin{align}
\label{eq1}
\boldsymbol y=\mathbf A\boldsymbol x,
\end{align}
where $K\leq M \ll L$, $\mathbf{A}$ is the $M\times L$ measurement matrix. Once the matrix $\mathbf{A}$ satisfies the restricted isometry property (\emph{RIP}) of order $K$ \cite{Candes05A}, \cite{Davenport11}, $\boldsymbol{x}$ can be uniquely restored from $M$ samples with reasonable sparsity-inducing constraint from noise-free measurements. In most real-world systems the measurements are likely to be contaminated by noise, however, it is still possible to reliabily recover the signal over a variety of noise models \cite{Candes06B}, \cite{Haupt06}. If the signal $\boldsymbol x$ is sparse, it can be recovered by
\begin{align}
\label{eq1_1}
\min_{\boldsymbol{x}}~\|\boldsymbol{x}\|_{0},~s.t.~ \boldsymbol{y}=\mathbf{A}\boldsymbol{x}.
\end{align}
The classical reconstruction algorithms include Orthogonal Matching Pursuit (OMP) \cite{Tropp07}, Subspace Pursuit (SP) \cite{Dai09}, $\ell_1$-magic algorithm in \cite{Kim07} where the minimization $\ell_0$-norm can be replaced by the minimization of $\ell_1$-norm, as well as Bayesian Compressed Sensing (BCS) \cite{Babacan10}.

\subsection{The Basic MIMO System}
We consider a typical MIMO system, as illustrated in Fig. 1, where the transmitter and the receiver is provisioned with $N_t$ and $N_r$ antenna, respectively. At the transmitter, after channel coding and the modulation according to the given constellation set $\mathcal{S}$, $M$ data streams will be transmitted simultaneously through $N_t$ transmit antenna, where $M= \min \{N_t, N_r\}$. 
We consider a flat fading MIMO channel, the received signal can be given by

\begin{figure}[t]
\centering {
\includegraphics [width=9cm]{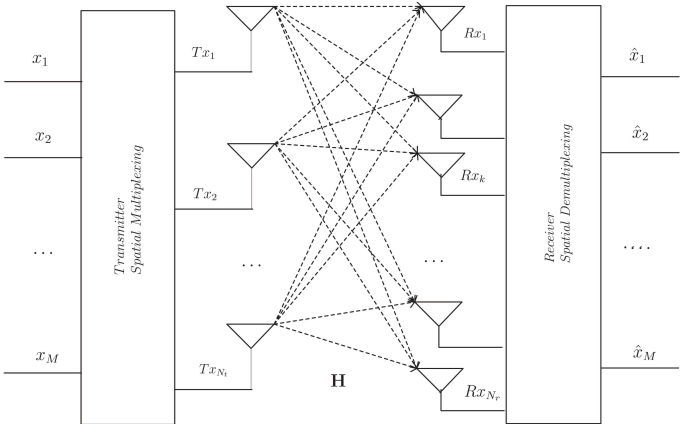} }
\caption{Illustration of the Spatial Multiplexing in MIMO system.}
\label{fig1}
\end{figure}

\begin{align}
\label{eq2}
\tilde{\boldsymbol y}=\tilde{\mathbf H} \tilde{\boldsymbol x}+\tilde{\boldsymbol v},
\end{align}
where $\tilde{\boldsymbol y}$ is the received signal of length $N_r$, $\tilde{\boldsymbol s}$ is the transmitted symbols vector with length $N_t$, $\tilde{\mathbf H}$ is the $N_r\times N_t$ channel matrix and $\tilde{\boldsymbol v}$ is complex normal zero-mean noise vector with covariance $\sigma^2 \mathbf I_{N_r}$. Each element of $\tilde{ \boldsymbol s}$ belongs to the constellation set $\mathcal{S}$. We assume the channel matrix $\tilde{\mathbf H}$ to be perfectly known at the receiver.

In order to avoid to handle complex-valued variables, let us define
\begin{align*}
\boldsymbol y=\begin{bmatrix}
    \mathfrak{\mathfrak{Re}\{\tilde{\boldsymbol y}\}}\\\mathfrak{\mathfrak{Im}\{\tilde{\boldsymbol y}\}}
  \end{bmatrix},
  \boldsymbol x=\begin{bmatrix}
    \mathfrak{\mathfrak{Re}\{\tilde{\boldsymbol x}\}}\\\mathfrak{\mathfrak{Im}\{\tilde{\boldsymbol x}\}}
  \end{bmatrix},
  \mathbf H=\begin{bmatrix}
    \mathfrak{\mathfrak{Re}\{\tilde{\mathbf H}\}} & -\mathfrak{\mathfrak{Im}\{\tilde{\mathbf H}\}} \\\mathfrak{\mathfrak{Im}\{\tilde{\mathbf H}\}} &\mathfrak{\mathfrak{Re}\{\tilde{\mathbf H}\}}
  \end{bmatrix},
\end{align*}
where $\mathfrak{Re}\{\cdot\}$ and $\mathfrak{Im}\{\cdot\}$ denotes the real part and imagniary part of a complex value signal. The size of $\boldsymbol y$ is $M=2N_r$, and the size of $\boldsymbol s$ is $N=2N_t$. The complex-valued signal model in \eqref{eq2} can be equivalently represented by the following real-valued model
\begin{align}
\label{eq3}
\boldsymbol y=\mathbf H\boldsymbol x+\boldsymbol v,
\end{align}
where $\boldsymbol v$ is defined as $\boldsymbol y$. In order to fully exploit the spatial multiplexing benefit of MIMO system, the precoding is widely employed. When linear precoding is assumed, the received signal can be given by
\begin{align}
\label{eq3_1}
\boldsymbol y=\mathbf {HP}\boldsymbol x+\boldsymbol v,
\end{align}
where $\mathbf P$ denotes the precoding matrix. Given the channel state information (CSI), the use of precoding matrix $\mathbf P$ at the transmitter side can be utilized to mitigate the inter-antenna interference and achieve the optimal power allocation among different spatial streams in terms of some optimization criteria. And our question here is, can we further improve the multiplexing gain of the MIMO system? And our motivation is to extend the idea of CS to the MIMO spatial multiplexing framework to derive the enhanced MIMO multiplexing gain.

\begin{figure}[t]
\centering {
\includegraphics [width=9cm]{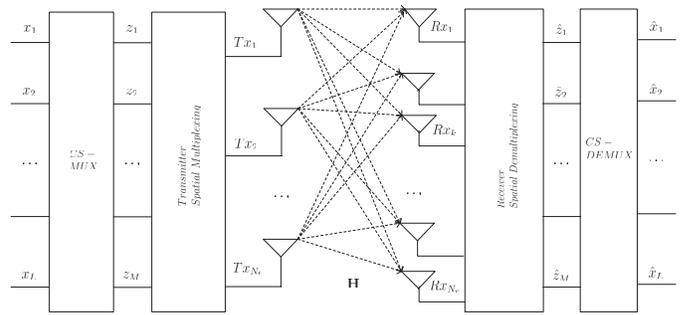} }
\caption{Illustration of the CS based Multiplexing for MIMO system.}
\label{fig1}
\end{figure}

\section{The CS-based MIMO Multiplexing Scheme}
In order to incorporate the CS into the MIMO system, we propose to introduce one additional CS multplexing module before the MIMO spatial multipelxing, as illustrated in Fig. 2. Accordingly, there will be an extra CS demultiplexing module after the MIMO spatial demultiplexing step ($i.e.$, the MIMO equalization). Now, instead of direct transmitting over the $N_t$ transmitter antenna, the modulated signal $\boldsymbol{x}$ will be firstly processed by using a measurement matrix $\mathbf \Phi$. The CS based multiplexing can be expliated as below
\begin{align}
\label{eq3_2}
\boldsymbol z=\mathbf \Phi \boldsymbol x,
\end{align}
where $\mathbf \Phi$ is the $M \times L$ multiplexing matrixㄛwhere $M=\rho L=min\{N_t, N_r\}$ and $\rho\in(0,1]$ represents the compression ratio. $\boldsymbol z\in \mathbb{C}^{M}$ represents the multiplexed vector. According to the principles of CS, multiplexing matrix $\mathbf \Phi$ must satisfy the RIP. And the received signal at the receiver side can be rewritten as
\begin{align}
\label{eq3_3}
\boldsymbol y=\mathbf H\boldsymbol z+\boldsymbol v=\mathbf {H\Phi} \boldsymbol x+\boldsymbol v.
\end{align}
%

\subsection{Overcomplete dictionary and Sparse Representation}
One may note from Eq. (7) that, it is an ill-conditioned problem to restore the transmitted signal $\boldsymbol x$ from the measurement $\boldsymbol y$, when $L>M$. According to the CS principle, in order to handle this issue, the proper over-complete dictionary of $\mathbf \Psi \in \mathbf{R}^{L \times d} $ and the sparse representation capability of the transmitted signal $\boldsymbol x$ over the dictionary $\mathbf \Psi$ will play an important role. Assume that $\boldsymbol x$ is sparse over the domain defined as $\mathbf \Psi$, that is to say, $\boldsymbol x=\mathbf \Psi \boldsymbol s$, where $\boldsymbol s$ is sparse, actually $\boldsymbol x$ and $\boldsymbol s$ are the same signal representation over different domain, then the most fundamental assumption of CS can be met.
There are so many regular dictionaries which are widely utilized in signal processing, and the examples include discrete cosine transformation basis, Fourier basis and Curvelets \cite{Candes04}. However, all these structured dictionaries can not well sparsely represent the signal. On the other hand, the learning based dictionaries were proposed to deduce the redundant dictionary from the signal samples. Compared to the regular structure dictionary, the learning based dictionary can better sparsely represent the natural signal. And the commonly used dictionary learning algorithms include the Method of Optimal Directions \cite{Engan99}, \emph{K-SVD} the (K-means Singular Value Decomposition) algorithm \cite{Aharon06}, the Iterative Subspace Identification algorithm \cite{Vikrham10}. Although the trained dictionaries is reasonable for natural analog signal, and the examples include image processing (such as image denoising and image super-resolution). However, there is few research report about the dictionary learning problem for digital signal, for instance the discrete modulation set $\mathcal{S}$, as we considered in this paper.

Obviously, an over-complete dictionary which subsumes all possible combinations of all the involved parallel modulation signal streams in $\boldsymbol x$
provides a heuristic method to derive the sparse representation $\boldsymbol{s}$ of $\boldsymbol{x}$, namely, $\boldsymbol{x}=\mathbf{\Psi} \boldsymbol{s}$.
Although this method may increase the computational complexity, it is still attractive in its potential to capture the sparse characteristics of the digital modulated signal $\boldsymbol{x}$. In order to reduce the realization complexity, we may consider to partition the transmitted signal $\boldsymbol{x}$ into sub-blocks,and we only need to calculate the exhaustive over-complete sub-dictionary. Now the received signal in Eq. (7) can be written as
%
\begin{align}
\label{eq5}
\boldsymbol y=\mathbf H \mathbf \Phi \boldsymbol x+\boldsymbol v= \mathbf H \mathbf \Phi \mathbf \Psi \boldsymbol s+\boldsymbol v.
\end{align}
Because the CS multiplexing matrix $\mathbf \Phi$ satisfies RIP and the sparse representation of the transmitted signal $\boldsymbol x$ over $\mathbf \Psi$ ($\|\boldsymbol s \|_0=1$ when the exhaustive dictionary is assumed), the CS based demultiplexing problem can be reformulated as the following standard CS signal reconstruction problem
\begin{align}
\label{eq7}
\min_{\boldsymbol{s}}~\|\boldsymbol{s}\|_{0},~s.t.~ \boldsymbol{y}=\mathbf{H\Phi\Psi}\boldsymbol{s}.
\end{align}

\subsection{The Unique Solution to the CS Demultiplexing Problem}
From the above analysis, it must be addressed that, now we can transmit $L$ modulated signal streams simultaneously through the MIMO system with $N_t$ transmit antenna and $N_r$ receive antenna. Because $L = {1\over \rho} \min\{N_t, N_r\}$, we now achieve the goal of improving the multiplexing gain of the MIMO system. Now we focus on how to ensure the unique restoration of the transmitted signal $\boldsymbol x$ from the measurement $\boldsymbol y$. Let $\mathbf{ \Upsilon=\Phi\Psi}$. Firstly, let's introduce a lemma.

\emph{Lemma 1. [Theorem 2.4 in \cite{Elad10}]} If a linear system of equations $\boldsymbol{z}=\mathbf \Upsilon \boldsymbol{x}$ has a solution $\boldsymbol{x}$ obeying $\|\boldsymbol x\|_0<spark(\Upsilon)/2$, this solution is necessarily the sparsest possible.

The \emph{spark} of a given matrix $\mathbf \Upsilon$ is the smallest number of columns from $\mathbf \Upsilon$ that are linearly-dependent. However, the \emph{spark} of a matrix is far more difficult to obtain, which calls for a combinatorial search over all possible subsets of columns from $\mathbf \Upsilon$. If $\mathbf \Phi$ comprises random $i.i.d.$ entries, for instance, Gaussian matrix as we assumed in the proposed CS based multiplexing scheme, then with probability 1 we have $Spark(\Phi)=M+1$, implying that no $M$ columns are linearly-dependent. Later, the new measure of quality \emph{RIP} of a matrix was introduced to replace \emph{spark}. The Gaussian matrix $\mathbf \Phi$ is utilized as the CS based multiplexing matrix in that it can satisfy the \emph{RIP} with overwhelming probability. 

\emph{Lemma 2. [Theorem 2.2 in \cite{Rauhut08}]} Let $\mathbf \Psi\in \mathbb{R}^{L\times d}$ be a dictionary of size $d$ in $\mathbb{R}^{L}$ with restricted isometry constant $\delta_{K}(\mathbf \Psi), K_0\in \mathbb{N}$. Let $\mathbf \Phi\in \mathbb{R}^{M\times L} (M=\rho L)$ be a random matrix satisfying RIP and assume $M\ge C\delta^{-2}\left(K\log(d/K)+\log(2e(1+12/\delta))+t \right)$
for some $\delta\in(0,1)$ and $t>0$. Then with probability at least $1-e^{-t}$ the composed matrix $\mathbf {\Upsilon=\Phi\Psi}$ has restricted isometry constant $\delta_{K}(\mathbf \Phi\Psi)\le \delta_{K}(\mathbf \Psi)+\delta\left(1+\delta_{K}(\mathbf \Psi)\right)$, the constant satisfies $C\le 9/c$ and $c>0$.

By the above lemma, we can readily derive that, the matrix $\mathbf \Upsilon$, which is a composition of a Gaussian random measurement matrix $\mathbf \Phi$ and a deterministic dictionary $\mathbf \Psi$, still has the related restricted isometry constants, which implies that $\mathbf {\Upsilon}$ still fulfills the \emph{RIP} requirement. So once there is sparsity characteristics in $\boldsymbol{s}$, we can obtain the unique sparsest solution $\hat{\boldsymbol s}$, and the restored transmitted signal can be reconstructed from $\hat{\boldsymbol x}=\mathbf \Psi\hat{\boldsymbol s}$. 

\subsection{The Practical CS based Multiplexing and Detection Design}
As we addressed before, when $L$ is large, the related over-complete dictionary $\mathbf \Psi$ has to consider all the possibilities of transmitted modulated signal from the given constellation set $\mathcal{S}$, which may make calculation complexity unaffordable. For the sake of practical system implementation, we proposed a sub-block based scheme. For the transmitted symbols vector $\boldsymbol x$ with length $L$, we divide it into $J$ group firstly, namely, $\boldsymbol x_j=[x_{j,1}; \cdots; x_{j,{L \over J}}], j=1, 2, \cdots, J$. For each group $\boldsymbol x_j$, we can have its CS multiplexing by $\boldsymbol z_j=\mathbf \Phi \boldsymbol x_j$. $\mathbf \Phi$ is the aforementioned Gaussian random measurement matrix with the dimension of ${M\over J} \times {L \over J}$. For each group, $\mathbf \Phi$ can be fixed and identical. By cascading all $J$ sub-blocks of $\boldsymbol z_j$ into one transmitted signal $\boldsymbol z$ we get the transmitted signal over $N_t$ transmit antennas. At the receiver side, we can try to restore each individual sub-block $\boldsymbol x_j$ based on the received signal sub-block $\boldsymbol y_j, j=1,\cdots, J$, where the size of the related dictionary $\mathbf \Psi$ is now $|\mathcal S|^{L \over J}$, where $|\mathcal S|$ denotes the size of the modulation alphabet $\mathcal S$. In this way, by constraining the sub-block size $J$, we can make the involved calculation complexity affordable in practical system realization. And the practical sub-block CS based multiplexing and detection procedure can be depicted by the pseudo-code in the pseudo-code of sub-block CS based multiplexing and detection algorithm.

-----------------------------------------------------------------------\\
\emph{The Pseudo-code of Sub-block CS based Multiplexing and Detection Algorithm}\\
$\mathbf{Transmitter}$: Given the transmitted signal $\boldsymbol x$, channel matrix $\mathbf H$ and the multiplexing matrix $\mathbf \Phi$;\\
(1) Divide the $L$ modulated signal stread $\boldsymbol x$ into $J$ sub-groups of $\boldsymbol x_j, j=1, 2, \cdots, J$;\\
(2) Multiplex each sub-group to get the CS-based multiplexing signal: $\boldsymbol z_j=\mathbf \Phi \boldsymbol x_j, j=1, 2, \cdots, J$;\\
(3) Cascade all $L$ multiplexing signal $\boldsymbol z_j$ into one vector $\boldsymbol z$ for transmission over all transmit antenna simultaneously; \\
$\mathbf{Receiver}$: Given the received signal $\boldsymbol y$, channel matrix $\mathbf H$, the multplexing matrix $\mathbf \Phi$ and dictionary $\mathbf \Psi$;\\
(1) Restore the transmited CS multiplexing signal $\hat{\boldsymbol z}$ via MIMO equalization;\\
(2) Divide $\hat{\boldsymbol z}$ into $J$ sub-groups $\hat{\boldsymbol z_j}, j=1, 2, \cdots, J$;\\
(3) For each group $\hat{\boldsymbol z_j}$, derive the sparse solution $\hat{\boldsymbol s}_j$ by solving $\min_{\boldsymbol{s_j}}~\|\boldsymbol{s}_j\|_{0},~s.t.~ \hat{\boldsymbol{z_j}}=\mathbf{\Phi\Psi}\boldsymbol{s_j}$; \\
(4) Reconstruct the $\hat{\boldsymbol x}_j$ as $\hat{\boldsymbol x}_j=\mathbf{\Psi} \hat{\boldsymbol s}_j$;\\
(5) Cascade all $\hat{\boldsymbol x_j}$ into the transmitted modulated signal $\hat{\boldsymbol x}$.\\
-----------------------------------------------------------------------\\

\section{Numerical Results}
In this section, we demonstrate the performance of the proposed CS-based MIMO multiplexing scheme. In all simulations, the QPSK modulation is assumed for simplicity. And it is assumed that all entries of the MIMO channel are independently and identically distributed, zero-mean, complex normal random variables (namely, Rayleigh fading is assumed). For each \emph{SNR}, we perform up to $100,000$ Monte Carlo simulations to obtain the average probability of error. 
As for the MIMO equaliation, we consider several other MIMO detectors. In the proposed CS-based MIMO multiplexing scheme, the zero-forcing (ZF) MIMO detector is assumed assumed. For comparison, we also include the ZF detector and SDR detector in \cite{Ma08} for the traditional MIMO system. Meanwhile, in all simulations, the LTE standardized $1/3$-code rate Turbo code is employed as the error control coding scheme. We assume the \emph{BCS} as the CS reconstruction algorithm \cite{Babacan10}. And $J=16$ sub-block size is assumed in the proposed CS based MIMO multiplexing scheme. Note that in all simulation figures, we use the notations of $(N_t, N_r)-M$ to explicate the MIMO and CS based multiplexing setup.

\begin{figure} \centering
\subfigure[The achieved sum rate] { 
\includegraphics[width=0.8\columnwidth]{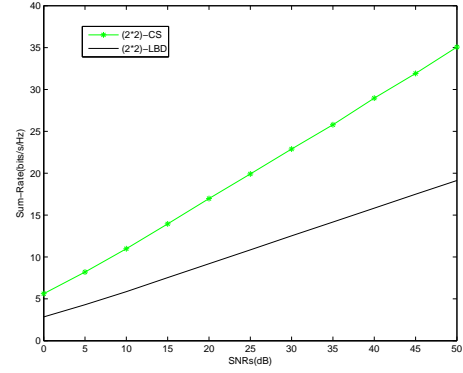}
}
\subfigure[The BER performance] { 
\includegraphics[width=0.8\columnwidth]{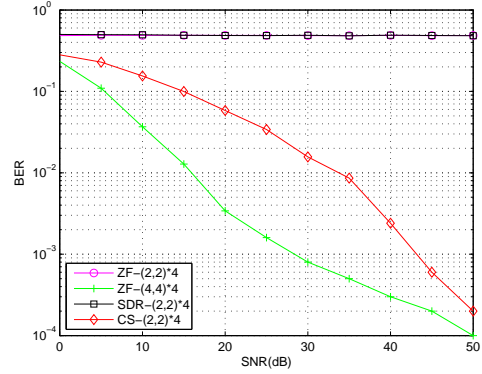}
}
\caption{$(N_t=2,N_r=2)$-MIMO System. }
\label{fig}
\end{figure}

\begin{figure} \centering
\subfigure[The achieved sum rate] { 
\includegraphics[width=0.8\columnwidth]{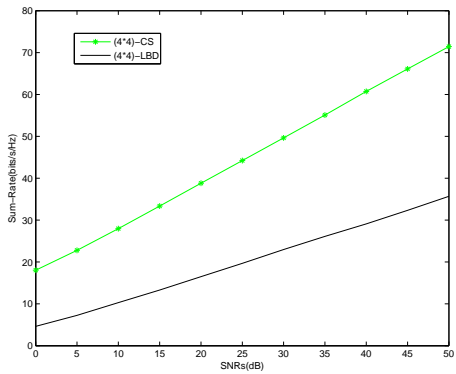}
}
\subfigure[The BER performance] { 
\includegraphics[width=0.8\columnwidth]{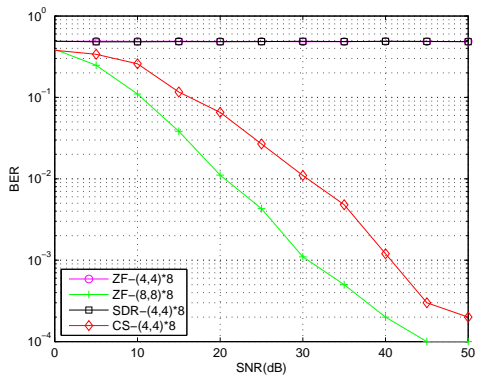}
}
\caption{$(N_t=4,N_r=4)$-MIMO System.}
\label{fig}
\end{figure}

And the achieved sum rate of the MIMO system, together with the achieved detection performance in terms of bit error rate are illustrated in Fig.3 - Fig.5 for $(2,2)-4, (4,4)-8$ and $(20,20)-40$ system setup, respectively. In sum rate calculations, the LBD scheme refers to the low-complexity block diagonalized-type precoding scheme in \cite{Zu13}.
One may readily observe from all simultation results that, in all cases, the proposed CS-based multiplexing can be employed to effectively improve the multiplexing gain. There is some detection reliability loss in the proposed CS-based enhanced MIMO multiplexing scheme, as illustrated in Fig.3-Fig. 5, however, the performance loss is acceptable, especially within moderate to high SNR region. In fact, as we know, the multiplexing gain is more desirable for MIMO system within high SNR region. And our proposed CS-based enhanced MIMO multiplexing can work well within the high SNR region as well. In the same time, we may also note that, without the use of our proposed CS-based multiplexing step to reduce the dimension of the transmitted modulated signal streams number to suit for the underlying spatial mulpelxing gain of the MIMO system, the direct increase in the number of simultaneous modulated signal streams to beyond the spatial multiplexing will totally fail in restore the transmitted signal, due to the fact that the detection problem is totally an ill-conditioned problem without the use of the proposed CS-based multiplexing mechanism to improve the existing MIMO multiplexing scheme.

\begin{figure} \centering
\subfigure[The achieved sum rate] { 
\includegraphics[width=0.8\columnwidth]{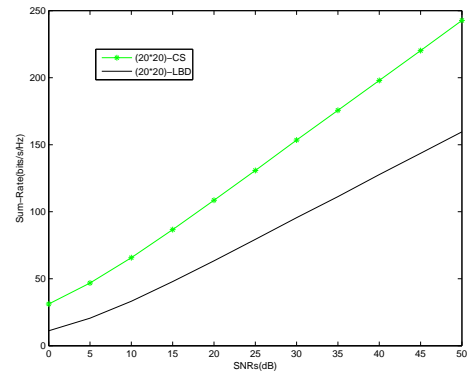}
}
\subfigure[The BER performance] { 
\includegraphics[width=0.8\columnwidth]{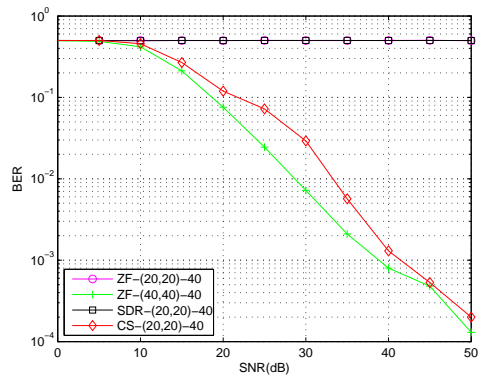}
}
\caption{$(N_t=20,N_r=20)$-MIMO System.}
\label{fig}
\end{figure}

%

\section{Concluding Remarks}
In this paper, we have proposed a new CS-based multiplexing approach to further improve the multiplexing gain of MIMO system. Our approach shows a considerable potential in broadband communication system, wherein higher transmission throughput is requested with limitd transmit and receive antenna in MIMO system. And the incurred complexity at the transmitter side is almost negligible since the proposed CS based multiplexing can be realized by using a Gaussian random matrix. By using the proposed sub-block processing, the detection complexity at the receiver can also be made feasible for practical system. It should be noted that, the proposed CS-based multiplexing scheme can be readily extended to the MU-MIMO, OFDM/OFDMA and other multiplexing scheme, which will be left for the next step analysis. Meanwhile, more efficient dictionary to enable sparse representation of multiple discrete modulation symbol streams is also another important issue to be explored in the next step work.

%






\end{document}